\documentclass{pasj00}
\usepackage{graphicx}
\usepackage{amsthm}

\begin{document}

\author{Fabiola M. A. Ribeiro \and Marcos P. Diaz}
\affil{Instituto de Astronomia, Geof\'{i}sica e Ci\^{e}ncias Atmosf\'{e}ricas,
Universidade de S\~{a}o Paulo,\\05508-900, S\~{a}o Paulo, SP, Brazil}
\email{fabiola@astro.iag.usp.br}
\title{A Time-Series Analysis of the H$\alpha$ Emission Line in V3885 Sagitarii}

\maketitle

\begin{abstract}
Flickering is a phenomenon related to the mass accretion observed among many classes of astrophysical objects. In this paper we present a study of the flickering emission lines and continuum of the Cataclysmic Variable V3885 Sgr. The flickering behavior is first analyzed through statistical analysis and lightcurves power spectra. Autocorrelation techniques are then employed to estimate the flickering flares timescales. A cross correlation study between the line and its underlying continuum variability is presented. The cross correlation between the photometric and spectroscopic data is also discussed. The periodograms, calculated using emission line data, show a behavior that is similar to those obtained from photometric datasets found in the literature, with a plateau at lower frequencies and a power law at higher frequencies. The power law index is consistent with stochastic events. The cross-correlation study indicates the presence of a correlation between the variability on H$\alpha$ and its underlying continuum. Flickering timescales derived from the photometric data were estimated as 25 minutes for two lightcurves and 10 minutes for one of them. The average timescales of the line flickering is 40 minutes, while for its underlying  continuum it drops to 20 minutes.
\end{abstract}

\KeyWords: accretion, accretion disks --- stars: binaries: close --- stars: cataclysmic variables

\section{Introduction}

Cataclysmic variables (CVs) are close binary systems composed of a white dwarf (the primary star) and a red dwarf or subgiant star (the secondary star). The secondary star fills its Roche lobe and matter is transfered to the primary, forming an accretion disk around the white dwarf. If the primary's magnetic field is strong enough, this accretion disk can be partially or totally disrupted and the accreted matter will follow the magnetic field lines through to the primary. One of the characteristics of the CVs is the presence of flickering.

Flickering is observed as stochastic variations in the emitted radiation. The flickering timescales range from seconds to several minutes and the amplitudes from hundredths of a magnitude to more than one magnitude. The first publications mentioning flickering on CVs were about T CrB \citep{Pet46}, AE Aqr \citep{Hen49} and WX UMa \citep{Lin49}. \citet{Gra55} verified that the flickering intensity on SS Cyg did not change during the outbursts. \citet{Pin59} showed that flickering is dependent on wavelength, revealing higher amplitudes at smaller wavelengths.

Flickering is not a peculiarity of the CVs, but a phenomenon associated with the mass accretion process itself. It is also observed among some symbiotic stars (e.g. CH Cyg, \cite{Mik90}), X-ray binaries \citep{Rey07, Mal03, Bap02} and pre-main sequence stars \citep{Ken00, Cla05}. However, it is also seen among systems where the accretion disk is absent, as magnetic CVs (e.g. AM Her, \cite{Bai77}).

Flickering can be originated from different regions of the binary system. \citet{War71} observed that the U Gem flickering disappeared during eclipses, and as the disk's central region was not occulted during those eclipses, they associated the flickering source with the hot-spot. Nonetheless, an opposite behavior was observed in OY Car \citep{Vog81}, HT Cas \citep{Pat81} and Z Cha \citep{Bru96}, since the flickering activity was observed even during hot spot eclipses. \citet{Hor85} associated the flickering source in eclipsing systems with the inner parts of the accretion disk. It was also associated with this same region in OY Car \citep{Hor94}. \citet{Bru96} concluded that the Z Cha flickering source was near the white dwarf, but also that other flickering sources could exist at other photometric states.

Many mechanisms are proposed to explain the flickering phenomenon in CVs. The first hypothesis, proposed by \citet{War71}, was that flickering is produced in the hot spot region. \citet{Sto79} suggested that condensations in the matter stream from the secondary to the primary are the flickering source in the polar AM Her. The mechanism proposed by \citet{Els82} is based on non-uniform mass accretion over the primary, due to instabilities at the inner accretion disk. Following this later model, white dwarf rotational modulation on the flickering should be noted, but this is not the case. Another flickering mechanism presented by \citet{Els82} was associated with turbulence in the accretion disk. In this model, a region of the accretion disk becomes turbulent, decaying to a lower potential disk radius, and the corresponding potential energy difference is released in the form of a flare. As the magnetic field of the disk is amplified by the dynamo effect, magnetic loops appear over the disk and the corresponding energy is freed as a flare when the magnetic loops decay. These models are discussed in detail by \citet{Bru89}.

Most of the works conducted on CVs flickering until now were based on the study of continuum flickering. As the lines seen in CVs have different ionization degrees (eg. H I, He I, He II) they should come from distinct regions of the system, so the study of flickering on spectral lines should provide complementary information about this phenomenon. The first line flickering study was on V442 Oph \citep{Dia01}, in which the flickering was mapped using tomographic techniques. Then another study was conducted on V3885 Sgr \citep{Rib07}. This phenomenon was also detected on the symbiotic star candidate AE Cir \citep{Men08}.


V3885 Sgr is one of the most luminous CVs known. It is a non eclipsing system, being classified as a UX UMa nova like star. Also, it has a high mass transfer rate \citep{Pue07}. This system was discovered by Bond \citep{Bid68}, showing a spectra with widened Balmer absorption lines. \citet{Bon71} identified the inclination of the continuum as similar to a DA white dwarf. \citet{Weg72} observed emission line cores superposed to the absorption lines, identifying the object as a nova-like CV. \citet{Hes72} first identified rapid variability on the system. The first estimate of the orbital period was of 0.2 days \citep{Cow77}. \citet{Hau85} revised this value to 0.259 days. From spectroscopic data in the blue, \citet{Har05} obtained a estimate of 0.207135 days for the orbital period, presented a radial velocity study and detected the illumination of the secondary star by the disk. \citet{Rib07} presented a long term ephemeris (with 0.20716071 days orbital period) for this system, emission line Doppler maps, and a mapping of the flickering sources using tomographic techniques. In the \citet{Rib07} work, the dominant flickering source detected was attributed to flickering from the disk being reprocessed in the illuminated face of the secondary star.

\section{Observations}

The spectrophotometric observations were conducted from 1999 to 2002, using the 1.6m telescope at Observatorio Pico dos Dias (LNA/CNPq) and the 1.5m telescope at Cerro Tololo Inter-American Observatory. A total of 1888 spectra with 2 \AA spectral resolution were obtained using differential spectrophotometric techniques. These observations and data reduction are described in detail in our previous work \citep{Rib07}.

Synthetic square band photometry at H$\alpha$ and its nearby continuum was performed from the observed spectra. The H$\alpha$ line flux was obtained by subtracting the flux at the H$\alpha$ region from the continuum. The ratio between the line and continuum integrated fluxes was also calculated. As the errors in flux calibration should affect identically both the line and continuum fluxes, its ratio should be absent of them. Synthetic lightcurves were calculated using these integrated fluxes, for line, continuum and line/continuum ratio.

The V3885 Sgr narrow band photometric data were obtained simultaneously with the spectrophotometric data at some observing nights (table 1). To obtain the photometric data in a blue band, free of spectral lines, the photometric observations were conducted using the Geneva B1 filter.

\begin{table}[htbp] 
\setlength{\belowcaptionskip}{10pt}
\caption{\it Simultaneous photometric observations.}
\begin{center}
\begin{tabular}{lccc}
\hline
date & telescope & \textit{time}\\
     &           & coverage\\
\hline
2001 Sept 10 & Zeiss 60 cm             & 1h 09m\\
2001 Sept 18 & Boller \& Chivens 60 cm & 4h 48m\\
2001 Sept 19 & Zeiss 60 cm             & 3h 22m\\
2002 July 07 & Boller \& Chivens 60 cm & 1h 38m\\
\hline \\
\label{table 1}
\end{tabular}
\end{center}
\end{table}

The reduction of the photometric data followed the standard procedures concerning bias and flatfield corrections.

As the photometric datasets were taken with different telescopes and detectors, the same field stars were not present on all the frames. Only one comparison star (RA 19 47' 33.7'' and dec -42 00' 32.5'', J2000) is common to all frames. This comparison star and the variable are marked on figure 1. We were unable to find any calibration of this star in the Geneva system in the available literature, so in this work we present photometric data as flux ratios.

\begin{figure}[htbp] 
\begin{center}
\FigureFile(80mm,80mm){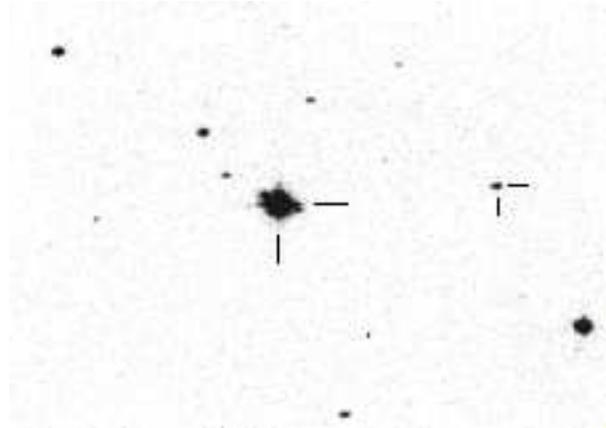}
\end{center}
\caption{Field observed from the 2001 September 10 night. The variable (at center) and the comparison star are marked.}
\label{fig 1}
\end{figure}

\textit{The photometric lightcurves are presented together with the ones obtained from H$\alpha$ synthetic photometry on figure 2}.

\begin{figure*}[htbp] 
\begin{center}
\rotatebox{-90}{%
\FigureFile(115mm,115mm){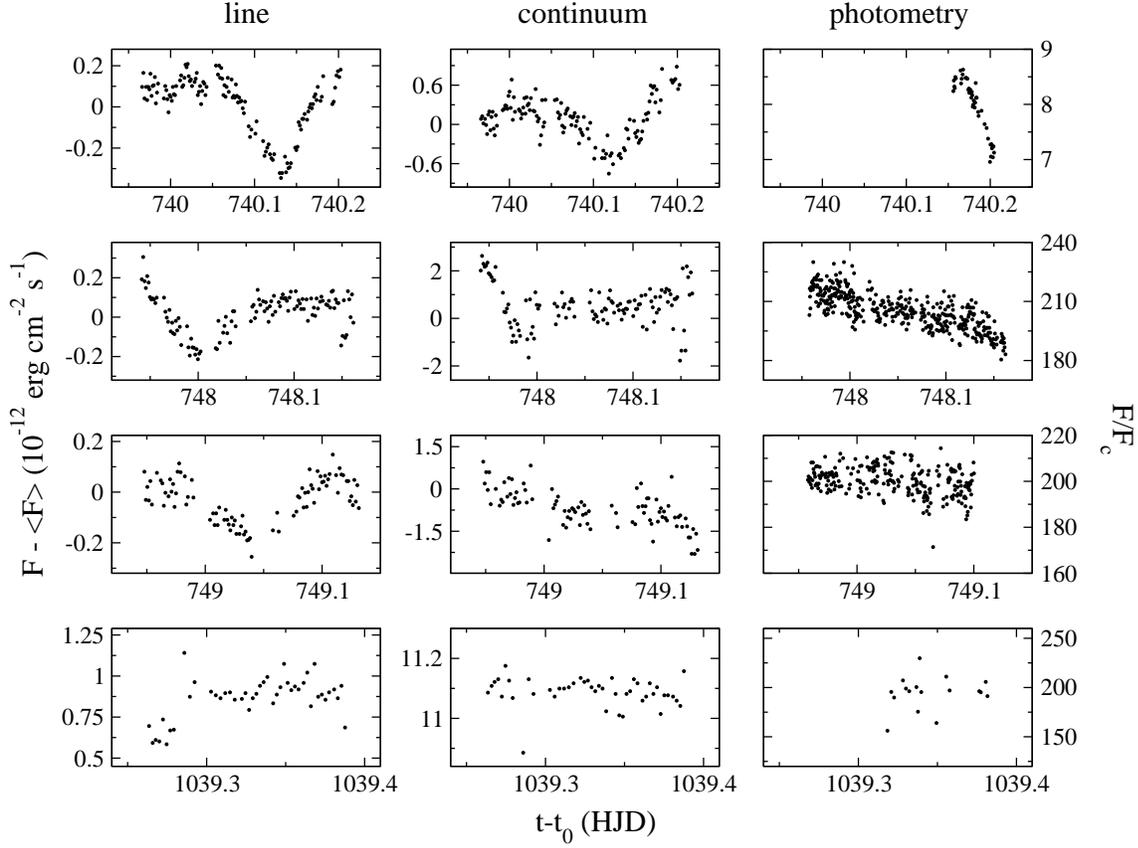}}
\end{center}
\caption{Synthetic lightcurves for H$\alpha$ line (first column), its continuum (second column) and the corresponding photometric lightcurves (third column). The observing nights are, from top to bottom, 2001 September 10, 2001 September 18, 2001 September 19 and 2002 July 07. F/Fc is the relative flux between the variable and the comparison star.}
\label{fig 2}
\end{figure*}

From figure 2 one can see that the photometric flux was smaller on the 2001 September 10 dataset than on the other observing nights. As the spectrophotometric data also presented a lower flux at this particular night, we can associate this behavior to a low state of the variable star.

\section{Results}

Some flickering parameters can be derived directly from the lightcurves. For each observing night, an estimate of the lightcurves variability can be obtained by calculating the standard deviation ($\sigma$) of the lightcurve points about their average. An estimate of the flickering amplitude ($A$) can be obtained from the difference between the maximum and minimum points of the lightcurve. To avoid the contamination of these quantities by spurious data points, the following procedure was adopted: a function was fitted to each night lightcurve to follow the global behavior of the points; the points beyond 3-$\sigma$ from this curve were rejected and this function was subtracted from the lightcurves. The result is a lightcurve with a ``flat'' global behavior. The activity was estimated as the standard deviation, and the amplitude as the difference between minimum and maximum points of this modified light curve. These values and their standard deviation are presented in table 2.

\begin{table*}[ht!] 
\setlength{\belowcaptionskip}{10pt}
\caption{\it Light curves parameters (in units of F$_\nu$= 10$^{-25}$ erg s$^{-1}$ cm$^{-2}$ for line and continuum).}
\begin{center}
\begin{tabular}{lcccccc}
\hline
& \multicolumn{2}{c}{line flickering}&\multicolumn{2}{c}{continuum flickering}&\multicolumn{2}{c}{line/continuum ratio}\\
\hline
date	     & $\sigma$ & A    & $\sigma$ & A    & $\sigma$ & A\\
\hline
1999 Sept 01 & 0.13     & 0.72 & 0.48     & 2.8  & 0.0051   & 0.029\\
1999 Sept 02 & 0.13     & 0.74 & 1.4      & 8.3  & 0.0040   & 0.025\\
2000 July 07 & 0.094    & 0.52 & 0.10     & 0.55 & 0.0048   & 0.027\\
2000 July 08 & 0.11     & 0.47 & 0.10     & 0.48 & 0.0051   & 0.022\\
2000 July 09 & 0.089    & 0.46 & 0.07     & 0.44 & 0.0053   & 0.024\\
2001 Mar 22  & 0.18     & 0.93 & 1.5      & 8.1  & 0.0072   & 0.035\\
2001 Sept 10 & 0.084    & 0.47 & 0.40     & 2.2  & 0.0033   & 0.018\\
2001 Sept 18 & 0.063    & 0.36 & 0.76     & 4.1  & 0.0020   & 0.011\\
2001 Sept 19 & 0.072    & 0.38 & 0.59     & 3.3  & 0.0027   & 0.013\\
2002 June 16 & 0.060    & 0.24 & 0.70     & 3.0  & 0.0019   & 0.010\\
2002 June 17 & 0.068    & 0.30 & 1.2      & 5.9  & 0.0022   & 0.010\\
2002 June 18 & 0.075    & 0.34 & 0.58     & 2.2  & 0.0043   & 0.016\\
2002 June 19 & 0.072    & 0.36 & 0.41     & 1.9  & 0.0056   & 0.027\\
2002 June 20 & 0.040    & 0.19 & 0.22     & 1.0  & 0.0039   & 0.015\\
2002 June 21 & 0.17     & 0.82 & 0.66     & 3.8  & 0.0045   & 0.022\\
2002 June 22 & 0.050    & 0.26 & 0.37     & 1.8  & 0.0018   & 0.011\\
2002 June 23 & 0.048    & 0.24 & 0.26     & 1.1  & 0.0035   & 0.016\\
2002 July 04 & 0.11     & 0.52 & 0.03     & 0.16 & 0.0029   & 0.012\\
2002 July 07 & 0.086    & 0.47 & 0.06     & 0.33 & 0.0035   & 0.010\\
\hline
mean values  & 0.46(5) & 0.091(9) & 2.7(6) & 0.52(10) & 0.0192(17) & 0.00388(33)\\
\hline
\label{table 2}
\end{tabular}
\end{center}
\end{table*}

\subsection{Flickering Frequency Study}

Periodograms were constructed from the synthetic H$\alpha$ lightcurves aiming to study the line flickering frequencies. The whole dataset was considered and the frequencies were constrained to have the orbital one as a lower limit. This procedure was repeated using data from the night with the largest temporal coverage. In both cases we could not identify periodic signals.

The periodograms were recalculated using a wider range of frequencies. These periodograms showed some peaks with significant power, but they can be attributed to the orbital period and its harmonics. A sinusoidal fit was then subtracted from the lightcurves as to remove any orbital feature from the periodogram. They are shown in a logarithmic scale on Figure 3.

\begin{figure*}[htbp]
\FigureFile(120mm,120mm){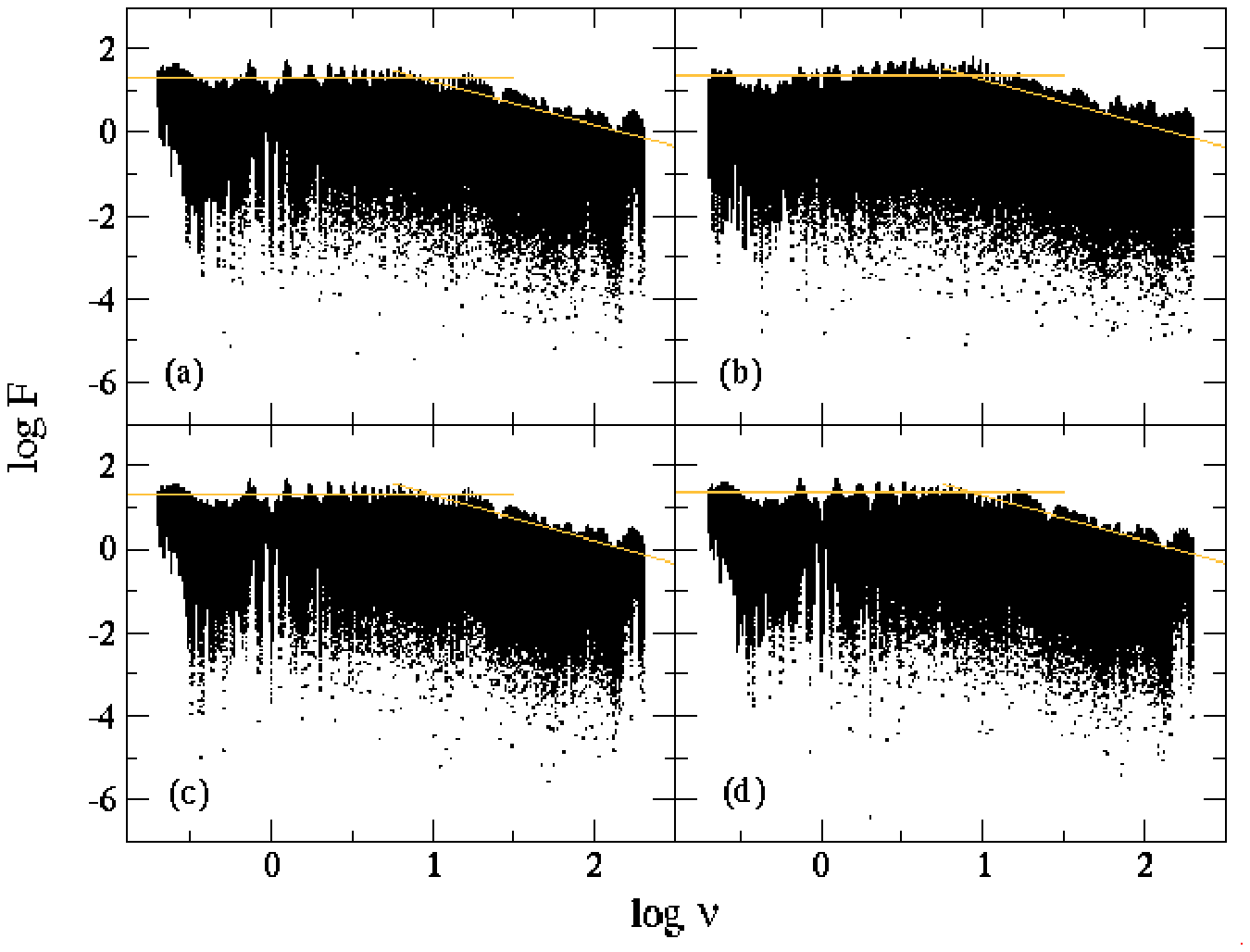}
\caption{Synthetic lightcurves periodograms from (a) H$\alpha$ line and continuum, (b) line/continuum ratio, (c) H$\alpha$ line only and (d) its underlying continuum. The periodograms show clearly two distinct behaviors: a plateau for lower frequencies and a descending region for higher frequencies. The lines are linear functions fitted to the plateau and descending graph regions in log-log scale.}
\label{fig 3}
\end{figure*}

Two distinct behaviors are seem on the periodograms of figure 3: they present a plateau at lower frequencies and a linear decay at higher frequencies. This behavior is also noticed on periodograms calculated from photometric observations (i.e. \cite{Kra99a, Kra99b}). A linear fit with zero slope was performed in the plateau region, while we used a linear least squares fit with arbitrary slope in the region with descending behavior. These two fitted lines were displaced by 0.3 in log(F) axis to follow the periodogram peaks. The slopes of the fit in the descending graph region are given on table 3. The uncertainties of the slope values were estimated only from the dispersion of the periodogram points and are just an indicative. A more fiducial determination should also take into account the propagation of the photometry flux uncertainties through the analysis procedure. The break frequency ($\nu_{break}$) was calculated for each case as the intersection of the plateau and descending fitted lines, its logarithmic value is also given on table 3.

\begin{table}[htb] 
\setlength{\belowcaptionskip}{10pt}
\caption{\it Power law function $F \propto f^\gamma$ indices adjusted to the periodograms.}
\begin{center}
\begin{tabular}{lcc}
\hline
periodogram          & $\gamma$ & log($\nu_{break}$) \\
\hline                                    
Continuum            & -1.10(4) & 1.0(3) \\ 
Line                 & -1.08(3) & 0.9(2) \\ 
Line + continuum     & -1.04(4) & 0.9(2) \\ 
Ratio line/continuum & -1.09(4) & 0.9(3) \\ 
\hline
\label{table 3}
\end{tabular}
\end{center}
\end{table}

Considering the uncertainties involved on this fit, we can conclude that the line, continuum, line+continuum and ratio periodograms follow a $F \propto f^{-1}$ power law, that is characteristic of stochastic events. \citet{Bru92} obtained $\gamma = -1.00(20)$ as an averaged index of the periodograms obtained from photometric observations of several CVs. \citet{Min94} relates the break frequency of the periodograms from black-hole objects to a critical radius, beyond which matter is accreted by viscous diffusion process, producing only white noise.


Periodograms were also calculated using the photometric lightcurves subtracted from the orbital component. The periodograms did not show peaks with power higher than about unity, as expected in the case of a periodic signal. We believe that photometric observations with larger temporal coverage should be necessary to perform this kind of frequency study.

\subsection{Flickering Autocorrelation}

The autocorrelation function always has a maximum at t=0. If the analyzed quantity is periodic, the autocorrelation function will exhibit a sequence of maxima and minima, as it measures how a quantity is related to itself at different time-lags. The flickering typical timescale can be estimated from the autocorrelation function width. As a degree of superposition in the flickering flares is expected, those timescales cannot be directly obtained from lightcurves.

The autocorrelation was calculated for each observing night for the H$\alpha$ line, continuum and line/continuum ratio synthetic lightcurves. They are presented on figures 4, 5 and 6. Those functions show two distinct behaviors: a narrow or a broad central peak. However, since the lightcurves were generated with a 0.002 days sampling and the narrow peaks have a similar timescale, these lightcurves are probably dominated by noise.

\begin{figure*}[htb!]
\FigureFile(145mm,145mm){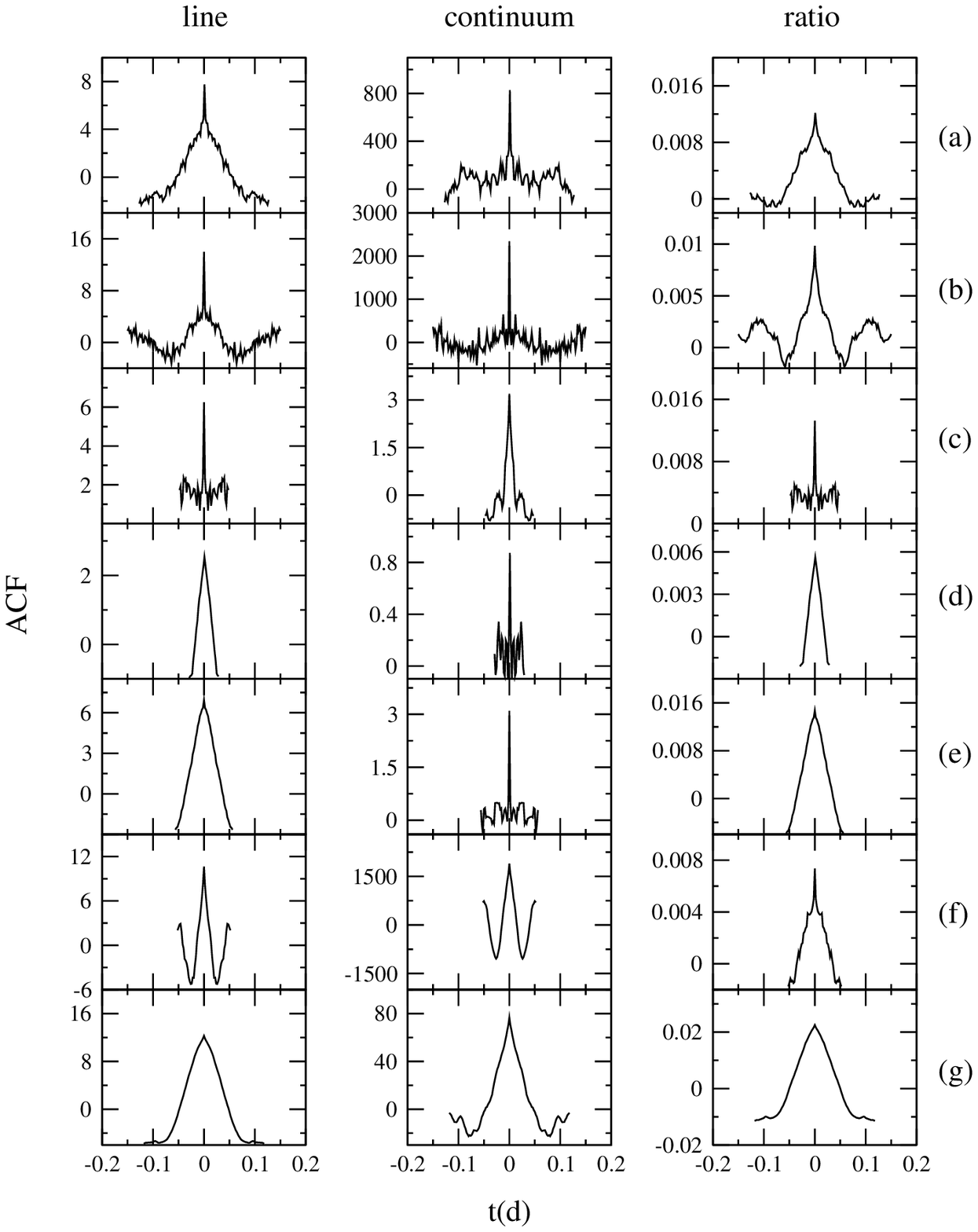}
\caption{Autocorrelation functions from the synthetic lightcurves, for each observing night: (a) 1999 Sept 01, (b) 1999 Sept 02, (c) 2000 July 07, (d) 2000 July 08, (e) 2000 July 09, (f) 2001 March 22 and (g) 2001 Sept 10.}
\label{fig 4}
\end{figure*}

\begin{figure*}[htb!]
\FigureFile(145mm,145mm){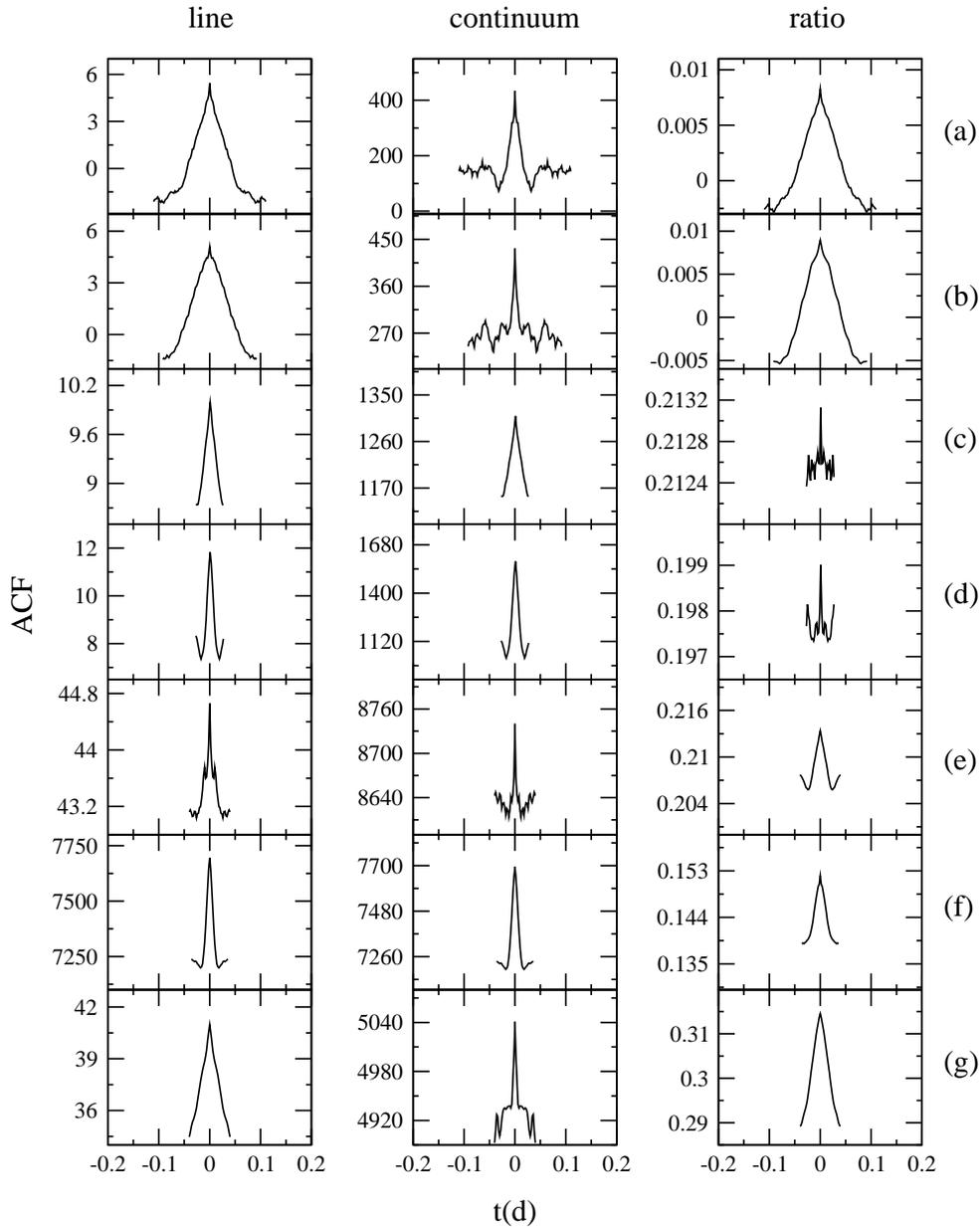}
\caption{Autocorrelation functions from the synthetic lightcurves, for each observing night: (a) 2001 Sept 18, (b) 2001 Sept 19, (c) 2002 June 16, (d) 2002 June 17, (e) 2002 June 18, (f) 2002 June 19 and (g) 2002 June 20.}
\label{fig 5}
\end{figure*}

\begin{figure*}[htb!]
\FigureFile(145mm,145mm){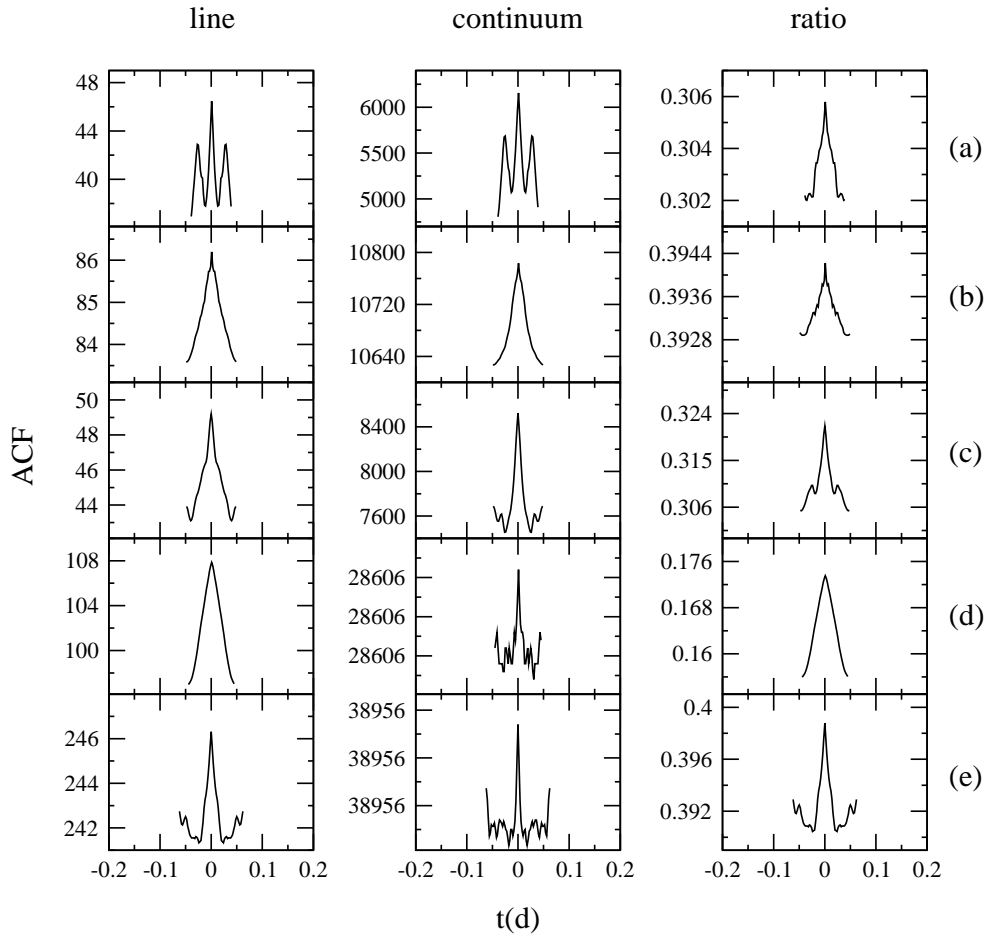}
\caption{Autocorrelation functions from the synthetic lightcurves, for each observing night: (a) 2002 June 21, (b) 2002 June 22, (c) 2002 June 23, (d) 2002 July 04 and (e) 2002 July 07.}
\label{fig 6}
\end{figure*}

\begin{table*}[htb] 
\setlength{\belowcaptionskip}{10pt}
\caption{\it FWHM of the flickering autocorrelation functions.}
\begin{center}
\begin{tabular}{lccccc}
\hline
& \multicolumn{3}{c}{synthetic lightcurves} & photometric lightcurves\\
& \multicolumn{3}{c}{FWHM (days)} & FWHM (days)\\
\hline
date	     & line flickering & continuum flickering & line/continuum ratio\\
\hline
1999 Sept 01 & 0.080  & 0.0034 & 0.069  &       \\
1999 Sept 02 & 0.060  & 0.0017 & 0.040  & -     \\
2000 July 07 & 0.0022 & 0.013  & 0.0021 & -     \\
2000 July 08 & 0.026  & 0.0027 & 0.026  & -     \\
2000 July 09 & 0.052  & 0.0021 & 0.053  & -     \\
2001 Mar 22  & 0.014  & 0.015  & 0.048  & -     \\
2001 Sept 10 & 0.074  & 0.055  & 0.076  & 0.018 \\
2001 Sept 18 & 0.062  & 0.026  & 0.072  & 0.016 \\
2001 Sept 19 & 0.072  & 0.0096 & 0.069  & 0.002 \\
2002 June 16 & 0.024  & 0.024  & 0.0015 & -     \\
2002 June 17 & 0.012  & 0.014  & 0.0027 & -     \\
2002 June 18 & 0.021  & 0.0035 & 0.018  & -     \\
2002 June 19 & 0.013  & 0.014  & 0.025  & -     \\
2002 June 20 & 0.038  & 0.058  & 0.035  & -     \\
2002 June 21 & 0.0061 & 0.011  & 0.028  & -     \\
2002 June 22 & 0.043  & 0.034  & 0.036  & -     \\
2002 June 23 & 0.039  & 0.018  & 0.015  & -     \\
2002 July 04 & 0.041  & 0.013  & 0.041  & -     \\
2002 July 07 & 0.020  & 0.0063 & 0.020  & 0.006 \\
\hline
$<$FWHM$>$  & 0.036(24) & 0.014(13) & 0.035(24) & 0.010(7)\\ 
\hline
\label{table 4}
\end{tabular}
\end{center}
\end{table*}

When the autocorrelation peak width is larger that the data sampling, one can consider that it is representative of the flickering timescale. The FWHM values for line, continuum and ratio are listed on table 4, including the autocorrelation peaks with widths similar to the sampling of the lightcurves. This calculation was repeated excluding the values where the FWHM was similar to the integration time, but the resulting values were very close to those obtained using all values.

The autocorrelation of the photometric data was also calculated (figure 7). The abscissa axis has the same scale for all graph panels. One can notice that the autocorrelation peak is larger for the (b) 2001 September 18 data, followed by (a) 2001 September 10 and narrower for (c) 2001 September 19 and (d) 2002 July 07 data. The FWHM of those peaks are given on the table 5.

\begin{figure}[htbp]
\begin{center}
\FigureFile(80mm,80mm){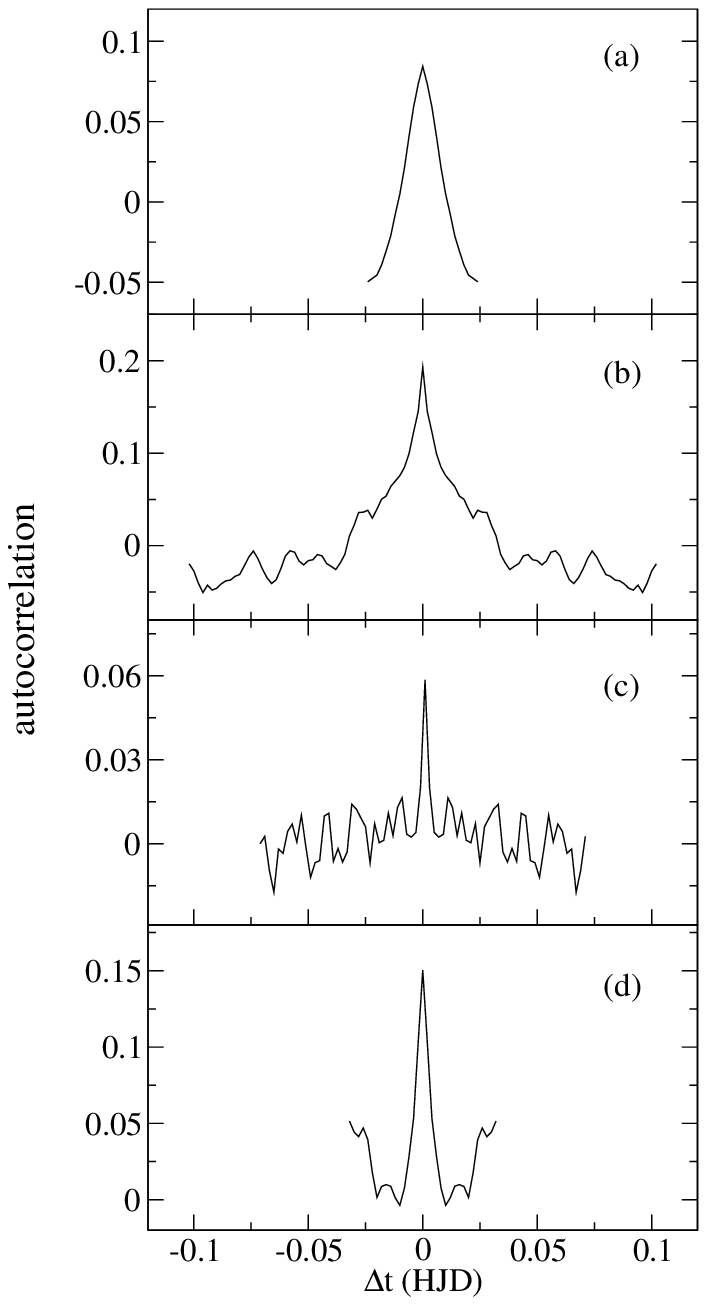}
\end{center}
\caption{Autocorrelation functions for the (a) 2001 September 10, (b) 2001 September 18, (c) 2001 Sept 19 and (d) 2002 July 07 lightcurves.}
\label{fig 7}
\end{figure}

By comparing the widths from photometric data presented in table 4 with the lightcurves sampling (0.002 days), one can see that the widths are bigger than this timescale for the two first nights and for the last one. On the night of September 19th the lightcurve presented a variability timescale of the same order as the lightcurve data sampling, thus the variability cannot be clearly detected (probably the variability has a timescale smaller that the data sampling). So, the detection of intrinsic variability is verified only on the lightcurves of 2001 September 10, 2001 September 18 and 2002 July 07.

\subsection{Flickering Cross-correlation}


Aiming to explore the line and continuum variabilities correlation, the cross correlation between line and continuum synthetic lightcurves was calculated. The cross-correlation quantifies how these two quantities are related at the same time or at different time lags. In our analysis, the peak of the line and continuum synthetic lightcurves cross-correlation function is displaced from the origin of 1.2 minutes. This delay cannot be explained by the light-time between the disk and the secondary as this later quantity has scale of seconds (more on the illumination hypothesis to explain the flickering from the secondary star, see \cite{Rib07}). A slight asymmetry on the cross-correlation profile can also be noticed. This analysis was repeated for each individual night. A clear correlation peak centered at t=0 was obtained only for data with longer temporal coverage.


We also used this technique to compare photometric and synthetic lightcurves. There is not a clear correlation peak for the three first nights, wince all the peaks have an autocorrelation intensity below 0.2, while values near to unity were expected if the correlation was significant. Nonetheless, on the June 2002 data, a clear correlation peak can be seen, with intensity of approximately 0.7 (0.74 for the correlation of the H$\alpha$ continuum and 0.64 for the H$\alpha$ line correlation with the photometric lightcurves). These peaks are displaced of -12 minutes for the continuum and 
23 minutes for the line. The peak widths are of about 10 minutes. As these correlations were calculated using photometry as a template and considering these autocorrelation peak displacements, the H$\alpha$ line lightcurves lags the B-band photometric lightcurve.

Aiming to verify if the correlation between the 2002 July 07 lightcurve and spectroscopic data is real or spurious, we superposed the photometric lightcurve with the synthetic ones, considering the displacement given by the autocorrelation peaks (12 minutes for the continuum and 23 minutes for the line). None correlation between the photometric and spectroscopic variability could be noticed from these graphs, indicating that the crosscorrelation peak observed was spurious, probably associated with the poor temporal coverage and the gaps in the photometric lightcurve.

\section{Conclusions}

A first study concerning the emission line flickering frequencies was presented. The periodograms calculated from the lightcurves showed a plateau behavior at lower frequencies and a power law decay at higher frequencies, as seen in the literature for other objects. We obtained a power law exponent of approximately -1 for the V3885 Sgr case. This value is characteristic of stochastic events.

There is a correlation between line and continuum flickering, with a single peak near zero time lag. This correlation was verified for the entire dataset as well as for the two nights of longest timeline.

An autocorrelation study was also performed for the individual observing nights. The FWHM of the autocorrelation function was used as an indicative of the flickering timescale. The mean flickering timescales are of about 40 minutes for the H$\alpha$ line and 20 minutes for its underlying continuum. The blue continuum flickering timescales are approximately 25 minutes for the first two photometry nights and 10 minutes for the last photometry night. The flickering variability could not be clearly detected on the third photometry night.

The flickering tomography study of other CV types is a perspective of future work. Flickering maps from lines of different ionization may also bring new information for the better understanding of flickering phenomenon. Data already taken to spectroscopic study of other CVs can be used to perform synthetic photometry and quantify the line flickering parameters on these objects.

\section*{Acknowledgments}
This work is based on data obtained at LNA/CNPq and Cerro Tololo observatories. F.M.A.R is grateful from support from FAPESP fellowship 01/07078-8 and 06/03308-2. MD acknowledges the support by CNPq under grant \#304043. We want to thank Alberto G.O.K. Martins for carefully reading the manuscript.


\begin{thebibliography}{}

\bibitem[Bailey, Mason \& Parkes(1977)]{Bai77}Bailey, J. A., Mason, K. O., Parkes, G. E., 1977, MNRAS, 180, 35
\bibitem[Baptista, Bortoletto \& Harlaftis(2002)]{Bap02}Baptista, R., Bortoletto, A., Harlaftis, E. T., 2002, MNRAS, 335, 665
\bibitem[Bidelman, MacConnell \& Bond(1968)]{Bid68}Bidelman, W. P., MacConnell, D. J., Bond, H. E., 1968, IAUC, 2085
\bibitem[Bond \& Landolt(1971)]{Bon71}Bond, H. E., Landolt, A. U., 1971, PASP, 83, 485
\bibitem[Bruch(1989)]{Bru89}Bruch, A., 1989, Habilitation Thesis, Westf\"{a}lische Wilhelms-Universit\"{a}t
\bibitem[Bruch(1992)]{Bru92}Bruch, A., 1992, A\&A, 266, 237
\bibitem[Bruch(1996)]{Bru96}Bruch, A., 1996, A\&A, 312, 97
\bibitem[Clarke et al.(2005)]{Cla05}Clarke, C., Lodato, G., Melnikov, S. Y., Ibrahimov, M. A., 2005, MNRAS, 361, 942
\bibitem[Cowley, Crampton \& Hesser(1977)]{Cow77}Cowley, A. P., Crampton, P., Hesser, J. E., 1977, ApJ 214, 471
\bibitem[Diaz(2001)]{Dia01}Diaz, M. P., 2001, ApJ, 553, 177L
\bibitem[Elsworth \& James(1982)]{Els82}Elsworth, Y. P., James, J. F., 1982, MNRAS, 198, 889
\bibitem[Haug \& Drechsel(1985)]{Hau85}Haug, K., Drechsel, H., 1985, A\&A, 151, 157
\bibitem[Henize(1949)]{Hen49}Henize, K. G., 1949, ApJ, 54, 89
\bibitem[Horne et al.(1994)]{Hor94}Horne K. et al., 1994, ApJ, 426, 294
\bibitem[Horne \& Stiening(1985)]{Hor85}Horne, K., Stiening, R. F., 1985, MNRAS, 216, 933
\bibitem[Grant(1955)]{Gra55}Grant, G., 1955, ApJ, 122, 566
\bibitem[Hartley et al.(2005)]{Har05}Hartley, L. E., Murray, J. R., Drew, J. E., Long, K. S., 2005, MNRAS, 363, 285
\bibitem[Hesser, Lasker \& Osmer(1972)]{Hes72}Hesser, J. E., Lasker, B. M., Osmer, P. S., 1972, ApJ, 176L, 31
\bibitem[Kenyon et al.(2000)]{Ken00}Kenyon, Scott J., Kolotilov, E. A., Ibragimov, M. A., Mattei, Janet A., 2000, ApJ, ,531, 1028
\bibitem[Kraicheva, Stanishev \& Genkov(1999)]{Kra99a}Kraicheva, Z., Stanishev, V., Genkov, V., 1999, A\&AS, 134, 263.
\bibitem[Kraicheva et al.(1999)]{Kra99b}Kraicheva, Z., Stanishev, V., Genkov, V., Iliev, L., 1999, A\&A, 351, 607.
\bibitem[Linnel(1949)]{Lin49}Linnell, A. P., 1949, Sky and Telescope, 8, 166
\bibitem[Malzac et al.(2003)]{Mal03}Malzac, J. et al., 2003, A\&A, 407, 335
\bibitem[Mennickent et al.(2008)]{Men08}Mennickent, R. et al., 2008, MNRAS, 383, 845
\bibitem[Mikolajewski, Mikolajewska \& Khudiakova(1990)]{Mik90}Mikolajewski, M., Mikolajewska, J., Khudiakova, T. N., 1990, A\&A, 235, 219
\bibitem[Mineshigem Ouchi \& Nishimori(1994)]{Min94}Mineshige, S., Ouchi, N. B., Nishimori, H., 1994, PASJ, 46, 97
\bibitem[Patterson(1981)]{Pat81}Patterson, J., 1981, ApJS, 45, 517
\bibitem[Pettit(1946)]{Pet46}Pettit, E., 1946, PASP, 58, 153
\bibitem[Pinto \& Rosino(1959)]{Pin59}Pinto, G., Rosino, L., 1959, Contr. Astrof. Asiago, 106
\bibitem[Puebla \& Diaz(2007)]{Pue07}Puebla, ........
\bibitem[Reynolds, Callanan \& Filippenko (2007)]{Rey07}Reynolds, M. T., Callanan, P. J., Filippenko, A. V., 2007, MNRAS, 374, 657
\bibitem[Ribeiro \& Diaz(2007)]{Rib07}Ribeiro, F. M. A., Diaz, M. P., 2007, AJ, 133, 2659.
\bibitem[Stockman \& Sargent(1979)]{Sto79}Stockman, H. S., Sargent, T. A., 1979, ApJ, 227, 197
\bibitem[Vogt, Kreminski \& Podersen(1981)]{Vog81}Vogt, N., Kreminski, W., Podersen, H., 1981, A\&A, 94L, 29
\bibitem[Warner \& Nather(1971)]{War71}Warner, B., Nather, R. E., 1971, MNRAS, 152, 219
\bibitem[Wegner(1972)]{Weg72}Wegner, G., 1972, PASAu, 2, 107

\end{thebibliography}
\end{document}